\documentstyle[prl,aps,epsf]{revtex}

\makeatletter
\def\preprint#1#2{%
\def\@preprint{
\noindent\raisebox{10mm}[0pt]{#1}\hfill\raisebox{10mm}[0pt]{#2}
\vskip -\baselineskip}%
}
\makeatother

\begin{document}
\draft
\preprint{}{ITP-UH-06/98}
\title{Spinon statistics in integrable spin-$S$ Heisenberg chains}
\vspace{1.5 em}

\author{Holger Frahm and Martin Stahlsmeier}
\address{Institut f\"ur Theoretische Physik, Universit\"at Hannover,
	 D-30167~Hannover, Germany}
\date{March 1998}
\maketitle
\begin{abstract}
  The spectrum of the integrable spin-$S$ Heisenberg chains is
  completely characterized in terms of spin-${1\over2}$ spinons.  In
  the continuum limit they form a quasi-particle basis to the higher
  level $SU(2)$ Wess-Zumino-Witten (WZW) models.  Enumerating the
  spinon states in finite systems we obtain effective single particle
  distribution functions for these objects which generalize Haldane's
  generalized exclusion principle to quasi-particles with non-Abelian
  exchange statistics.
\end{abstract}

Haldane's notion of fractional exclusion statistics \cite{hald:91} has
been used successfully to characterize elementary excitations in low
dimensional quantum systems.
For the excitations in the lowest Landau level arising in the
fractional quantum Hall effect (FQHE) this concept has been shown to
coincide with the usual definition of fractional statistics, namely
through their properties under exchange of anyons
\cite{wilczek:anyons}.  A complete description in terms of quasi
particles obeying fractional statistics is possible for integrable
quantum systems where detailed knowledge of the spectrum allow for an
classification of all states in terms of the elementary excitations.
Their statistical properties are described in terms of the statistical
interaction parameters $g_{\alpha\beta}$ which determines the many
particle Hilbert space dimension
\begin{equation}
  W = \prod_{\alpha} {d_\alpha + N_\alpha -1 \choose N_\alpha}\ .
\end{equation}
Here 
\begin{equation}
  d_\alpha=d_\alpha^{(1)} - \sum_\beta g_{\alpha\beta} (N_\beta -
  \delta_{\alpha\beta})
\label{statint}
\end{equation}
is the dimension of the Hilbert space of a single particle of species
$\alpha$ in the presence of $N_{\alpha}-1$ other particles of that
species at fixed coordinates.  $d_\alpha^{(1)}$ are constants which
can be interpreted as the dimensions of the single particle Hilbert
spaces.  This definition includes bosons ($g_{\alpha\beta}\equiv 0$)
and fermions ($g_{\alpha\beta}= \delta_{\alpha\beta}$).  An new
feature in this picture are the possible off diagonal statistics
parameters $g_{\alpha\ne\beta}$ encoding the mutual statistical
interaction of different species of anyons.  This is realized in
spin-${1\over2}$ Heisenberg chains \cite{hald:91} and has recently
been discussed in the context of the statistics of the quasiparticles
and -holes in the FQHE at filling $\nu=1/m$ with $m$ an odd integer
\cite{suwy:96,iscj:96}.

The approach outlined above allows to introduce single particle
distribution functions which already contain all information on the
statistical properties of the elementary excitations of the system
thereby giving an easy access to thermodynamic properties of the
system such as the low temperature specific heat.  Unfortunately, it
is easily seen that Haldane statistics (\ref{statint}) fails to
describe correctly the interesting case of quasiparticles obeying
non-Abelian statistics as realized for example in the FQHE at half
integer fillings \cite{more:91,blwe:92}: the distribution functions
obtained within this picture always lead to a low temperature specific
heat characteristic of a Gaussian conformal field theory (CFT) with
central charge $c=1$ \cite{iamp:96} while the quasi particles for the
$\nu={1\over2}$ FQHE are described by a $SU(2)_2$ WZW model with
$c=\frac{3}{2}$.

Recently, an extension of Haldane statistics to quasi particles
obeying non-Abelian exchange statistics has been proposed based on
recursion relations for truncated conformal spectra
\cite{scho:97,scho:98}: constructing the basis of the CFT explicitely
in terms of quasi-particle operators simple distribution functions
could be calculated that coincide with the ones found from
(\ref{statint}) for Abelian statistics and correctly reproduce the
central charge for the non-Abelian case.

In this letter we apply this method to characterize the spinon
excitations of the Bethe Ansatz soluble spin-$S$ $SU(2)$ Heisenberg
chains \cite{takh:82,babu}.  The particular form of the Hamiltonian
is not of interest here, it contains exchange interactions of nearest
neighbour spins only that are given as polynomials of degree $2S$ in
$\mathbf{S}_j\cdot\mathbf{S}_{j+1}$.
The eigenstates for the spin $S$ chain of length $L$ are characterized by
complex numbers $\lambda_j$, $j=1,\ldots,M$ satisfying the system of
Bethe Ansatz equations
\begin{equation}
  \left(  {{\lambda_j +iS}\over{\lambda_j-iS}} \right)^L
     =  \prod_{k\ne j}
        {{\lambda_j-\lambda_k+i}\over{\lambda_j-\lambda_k-i}}\ .
\label{bae}
\end{equation}
It is well known that regular Bethe states (corresponding to solutions
with $\lambda_j<\infty$) are highest weight states with spin $S_{tot}
= LS-M$, limiting $M/L$ by $S$ \cite{fata:84}. Moreover, in the
thermodynamic limit $L\to\infty$ all solutions of (\ref{bae}) are built
from so called strings $\lambda^{(n)}+{i\over2}\left(n+1-2j\right)$
with real centers $\lambda^{(n)}$ and $j=1,\ldots,n$.

Based on this classification the completeness of Bethe states has been
proven for the periodic chain \cite{kiri:83}: after rewriting (\ref{bae})
in terms of the real $\lambda^{(n)}$ any Bethe state is characterized
uniquely by a collection of integer or half integer numbers $Q_j^{(n)}$.
Now the number of possible states built from $\nu_n$ $n$-strings is just
the number of admissible choices of quantum numbers $Q_j^{(n)}\ne
Q_k^{(n)}$ for $j\ne k$.  The latter are found to be restricted to the
intervals $[-Q_{\max}^{(n)},Q_{\max}^{(n)}]$ where
\begin{equation}
  2 Q_{\max}^{(n)} +1 = L\min(n,2S) +\nu_n - \sum_{m=1}^\infty G_{mn} \nu_m\ 
\label{counter}
\end{equation}
with $G_{mn} = 2\min(m,n)$.  In particular, the antiferromagnetic
vacuum of a chain of even length is the {\em unique} singlet state in
which all $L/2$ modes for the $2S$-strings are occupied.

In a recent study \cite{fuky:96} of a related {\em ideal} system the
``strings'' appearing as solutions of a simplified version of the
Bethe Ansatz equations (\ref{bae}) have been interpreted as the
elementary objects spanning the Fock space of the system.  In this
picture one obtains the infinite dimensional matrix $G_{mn}$ for the
statistical interaction of the system (see also \cite{bewu:94}).
One should note, however,
that the role of the various strings in building the excitations over
the \emph{physical} vacuum, namely the antiferromagnetic ground state
of the system is quite different: in the thermodynamic limit, the
energy and momentum of a configuration is completely determined by the
distribution of holes in the sea of $2S$ strings, while the ones with
$n\ne 2S$ are merely responsible for the correct counting of states
\cite{takh:82}.  In particular, different configurations of the
$n>2S$-strings produce the $2^m$-fold degeneracy of the multiplet due
to the spin ${{1\over2}}$ degree of freedom carried by the $m$ holes
(spinons) while the $n<2S$ strings distinguish states
with the same number of holes and total spin.  Their presence is
reflected in the $S$-dependence of the critical behaviour of the
system \cite{babu}: the low-energy effective field theory of the
system has been identified as a $SU(2)_{2S}$ WZW model.  Separating
this field theory into a bosonic sector (with central charge $c=1$)
and a $Z_{2S}$ parafermionic sector \cite{tsve:88} the
$n<2S$ strings have been assigned to the latter.  In a different
approach they have been identified as spanning the
state space of a restricted solid on solid (RSOS) model on the basis
of the number of $m$-particle states \cite{resh:91}.

First let us briefly recall the situation in the $S={{1\over2}}$ case
where the ground state is a completely filled band of $\nu_1=L/2$
(real) 1-strings.  The elementary excitations above the
antiferromagnetic vacuum can be identified with holes in their
distribution \cite{fata:84}.  For chains of even (odd) length $L$
states with even (odd) number $m$ of holes form ``Yangian'' multiplets
containing states with total spin $S=0({1\over2}), \ldots, {m/2}$
which differ in the number of $n>1$-strings fixing the total
spin of the resulting state.  Counting these states (including their
$SU(2)$ multiplicity) the dimension of the $m$-spinon subspace is
\begin{equation}
   d_m = {L+1\choose m}\ .
\label{dim_sp2}
\end{equation}
Summing over the admissible $m$ this gives the correct total number of
states $2^L$.  Here the statistics of the excitations is modeled by
hard core lattice bosons reflecting the level--1 $SU(2)$ Kac-Moody
symmetry of the critical theory.
Taking into account the spin degree of freedom one finds that the
spectrum of the $m$-spinon states decomposes exactly into the states
present in the tensor product of $m$ spins ${{1\over2}}$: in the $m=2$
spinon sector there are ${{{L/2}+1}\choose2}$ triplets (built from
1-strings only) and ${L/2\choose2}$ singlets (containing a single
2-string).  Thus we are led to distinguish spinons with different spin
projection and to rewrite (\ref{dim_sp2}) as a sum over configurations
with $n_\uparrow$ ($n_\downarrow=m-n_\uparrow$) spinons with spin
$\uparrow$ ($\downarrow$) as
\[
  d_m = \sum_{\{n_\sigma\}} \delta_{n_\uparrow+n_\downarrow,m}
   \prod_{\sigma} \left( \begin{array}{c} d_\sigma+n_\sigma-1\\ 
                       n_\sigma\end{array}\right)
\]
where $d_{\sigma}(n_\uparrow,n_\downarrow) = ({{1\over2}})
\left(L-n_\uparrow -n_\downarrow\right) +1$.  This gives
$g_{\sigma\sigma'}={{1\over2}}$ establishing the semionic nature of
the spinons just as in the ideal spinon gas realized in the
Haldane-Shastry model \cite{hald:91,HaldShas}.  The difference between
this model and the nearest neighbour Heisenberg model considered here
is that a spinon-spinon interaction lifts the degeneracy of states
with different total spin in the super multiplets when their density
becomes finite.  Following Ref.~\cite{isak:94} the distribution
function of spinons is
\begin{equation}
  n(\epsilon) = 
  \frac{2}{1+{\rm e}^{\beta(\epsilon-\mu)}}\
\label{dist1}
\end{equation}
which gives the low temperature specific heat $C/L=\pi T/3v_F$ of a
$c=1$ model ($v_F$ is the Fermi velocity of the quasi particles).

For $S>{{1\over2}}$ one finds from (\ref{counter}) that there is a one to one
mapping between the possible configurations of $n\ge 2S$ strings between
the spin-$S$ model and the spin-${{1\over2}}$ Heisenberg chain (see
Table~\ref{tab:no_conf}):
for any $m$-spinon Bethe state with total spin $\le {m/2}$ in
the $S={{1\over2}}$ chain with string configuration
$\{\nu_n^{(1/2)}\}$ there exist Bethe states in the spin-$S$ chain
with string configuration $\{\nu_n^{(S)}\}$ such that
$\nu^{(S)}_{2S+n-1} = \nu_n^{(1/2)}$ for $n>1$ with the same number
$m$ of holes in the distribution of $2S$-strings.  In addition,
there are certain $n<2S$ strings present in these
states.  Note that this is seen only when considering holes,
the number $\nu_{2S}$ of $2S$-strings differs for
different configurations of the shorter strings.
Proceeding as in the $S={1\over2}$ case the number of $2m$ spinon states in
a chain of even length $L$ is found to be
\begin{eqnarray}
  d_{2m} &=&  \sum_{n_1=0}^{\left[ (\frac{2S-1}{2S}) m \right] }
  \sum_{n_2=0}^{\left[ (\frac{2S-2}{2S-1})n_1 \right]} \ldots
  \sum_{n_{2S-1}=0}^{\left[\frac{n_{2S-2}}{2} \right]}
  \left[ {L+1+2n_1 \choose 2m}
      \prod_{i=0}^{2S-2} {n_i+n_{i+2} \choose 2n_{i+1}} \right] 
\label{spinonS}
\end{eqnarray}
with $n_0\equiv m$, $n_{2S}\equiv0$ in the product.
Eq.~(\ref{spinonS}) and a similar expression for odd $L$ correctly
reproduce the total number of states for the spin-$S$ chain.  However,
an interpretation in terms of Haldane statistics, i.e.\ the ansatz
(\ref{statint}) for spin-${1\over2}$ quasi particles with an additional
internal degree of freedom turns out to be impossible.

To proceed we apply the method introduced in \cite{scho:97,scho:pc}.  
We identify the Bethe states of the spin chain with the spinon basis
of the chiral $SU(2)_{k=2S}$ WZW model proposed in \cite{bolk:95}
(a realization of the chiral CFT is the open boundary
version of the Takhtajan-Babujian models whose eigenstates are
enumerated by the same relation (\ref{counter})).  In this basis the
$m$-spinon states are of the form
\begin{equation}
  \phi^{\sigma_m}{{1\over2}\choose j_m\,j_{m-1}}_{-\Delta_m-n_m}\cdots
  \phi^{\sigma_2}{{1\over2}\choose j_2\,j_1}_{-\Delta_2-n_2}
  \phi^{\sigma_1}{{1\over2}\choose j_1\,0}_{-\Delta_1-n_1}
  |0\rangle\
\label{spbasis}
\end{equation}
where each operator $\phi^{\sigma_i}_{-\Delta_i-n_i}$ creates a spinon
with spin projection $\sigma_i=\uparrow,\downarrow$ and energy
$\epsilon(n_i+\Delta_i) = (2\pi v_F/L) (n_i+\Delta_i)$ ( $|0\rangle$
is the Fock vacuum).  The numbers $j_i$ in (\ref{spbasis}) take values
$0,{1\over2},\ldots,{k\over2}$ subject to the fusion rules
$|j_i-j_{i-1}|={1\over2}$ of the CFT and we have defined
$\Delta_i=\Delta(j_i) -\Delta(j_{n-1})$ with $\Delta(j)=j(j+1)/(k+2)$.
A basis of the spinon Fock space is provided by the states
(\ref{spbasis}) with $\sigma_1 =\ldots=\sigma_{N_\uparrow}=\uparrow$,
$\sigma_{N_\uparrow+1}=\ldots
=\sigma_{N_\uparrow+N_\downarrow}=\downarrow$ with the modes
$n_i\equiv n_{i,\min} + \tilde{n}_i$ and non decreasing sequences of
non negative integers $\left\{ \tilde{n}_i
\right\}_{i=1}^{N_\uparrow}$ and $\left\{ \tilde{n}_i \right\}_{i=
N_\uparrow+1}^{N_\uparrow+N_\downarrow}$.  The minimal allowed mode
sequence is constructed from $n_{1,\min}=0$ and
\[
  n_{i+1,\min} = \left\{
  \begin{array}{ll}
  n_{i,\min} + 1	& {\rm if~} j_{i+1}=j_{i-1}<j_i \\
  n_{i,\min}	& {\rm otherwise}
  \end{array} \right.
\]

To compute the distribution function for these modes we
introduce truncated partition functions
$q^{\Delta(j)} X_\ell^{(j)}(q,x,h)$ of the chiral CFT for which the
occupied spinon modes should satisfy $n_N\le\ell$ and $j_N=j$
\cite{scho:97,scho:pc}.  Here
$\mu+\sigma h$ are chemical potentials for the spinons
and we write $q={\rm e}^{-\beta(2\pi v_F/L)}$, $x={\rm e}^{\beta\mu}$.
These partition functions satisfy recursion relations
\begin{equation}
 \left( \begin{array}{c}
	X_\ell^{(0)}\\
	\vdots\\
	X_\ell^{(k)}
	\end{array} \right) =
 {\cal R}_\ell(q,x,\beta h)
 \left( \begin{array}{c}
	X_{\ell-1}^{(0)}\\
	\vdots\\
	X_{\ell-1}^{(k)}
	\end{array} \right)
\label{rec}
\end{equation}
with initial condition $X_0^{(j)}=c_jx^j$, $c_j=\sinh((j+1)\beta h/2)/
\sinh(\beta h/2)$.  The matrix ${\cal R}_\ell = R(xq^\ell,\beta
h)\,D(x^2q^{2\ell-1})$ is given in terms of $D(y)={\rm
diag}(1-{y},\ldots,1-{y},1)$ and the symmetric matrix
\begin{equation}
  R(y,\beta h) = \left( \begin{array}{ccccc}
1& c_1 y      & c_2y^2        & \ldots & c_{2S}y^{2S} \\
 & (1+c_2y^2) & (c_1y+c_3y^3) & \ldots & c_{2S-1}y^{2S-1}+c_{2S+1}y^{2S+1} \\ 
 &            & (1+c_2y^2+c_4y^4) & \ldots & c_{2S-2}y^{2S-2}+c_{2S}
	y^{2S}+c_{2S+2}y^{2S+2} \\
 & { \bf *}   &                   & \ddots & \vdots \\
 &            &                   &        & \sum_{i=0}^{2S}c_{2i}y^{2i}
		  \end{array}\right)\ .
\end{equation}
%
Below it will be useful to factorize the matrix ${\cal R}_\ell$ in
(\ref{rec}) further as $\bar{\cal R}_\ell \bar{\cal
R}_{\ell-{1\over2}}$ with $\bar{\cal R}_\ell = \bar{R}(xq^\ell, \beta
h) D(x^2q^{2\ell-1})$.  The non-zero matrix elements of this new
matrix are $\left(\bar R(y,\beta h)\right)_{i,k-j}=c_{i-j}y^{i-j}$
for $0\le i-j \le k$.

Remarkably, the number of $m$-spinon states contributing to the
partition function $X_\ell^{(0)}$ coincides with (\ref{spinonS}) for
the spin-$S$ chain of even length $L=2\ell$.  Since the finite size
spectrum of low lying excitations of the spin chain is determined by
the operator content of the $SU(2)_k$ WZW model we can identify
$X_\ell^{(0)}$ with the partition function ${\cal Z}_L$ of the spin
chain of length $L=2\ell$ and similarly $q^{\Delta(k)}X_\ell^{(k)}
={\cal Z}_{2\ell-1}$.  In this sense the partition function of the
integrable Heisenberg chain can be generated by a recursion relation
similar to (\ref{rec}) with the matrix $\bar{\cal R}_\ell$.  In
addition, this observation provides further evidence for the
equivalence of the spinons of the integrable chain with the ones
forming the quasi particle basis (\ref{spbasis}) to higher level
$SU(2)$ WZW models of Ref.~\cite{bolk:95}.

Viewing the $k+1$ spinon modes added in the $\ell^{\rm th}$ iteration
of (\ref{rec}) as a single degenerate level in the single particle
spectrum we can now approximate the partition function of the spin
chain as ${\cal Z}_{2\ell} = \prod_{i=1}^\ell \lambda_i^2$ with the
maximal eigenvalues $\lambda_i=\lambda^{(+)}(y=xq^i,\beta h)$ of the
recursion matrix $\bar{R}(y,\beta h) D(y^2)$.
This leads to a spinon distribution function
\begin{equation}
  n(\epsilon) = 2y\partial_y \ln \lambda^{(+)}(y,\beta h)\ , \quad
		y={\rm e}^{-\beta(\epsilon-\mu)}
\label{distk}
\end{equation}
for the spin-$S$ chain.  For $k=1$ this function has been studied in
Ref.~\cite{scho:97} where complete agreement with the corresponding
results from Haldane exclusion statistics (e.g.\ (\ref{dist1}) for
$h=0$) could be established.
In the general non-Abelian case $k=2S>1$ it is not possible to obtain
a closed expression for $n(\epsilon)$, for large negative $\epsilon$
we find $n(\epsilon)\sim 2k -(2+4\delta_{k,2}) {\rm
e}^{2\beta(\epsilon-\mu)}$.  The maximal occupation $4S$ of a level
containing $2S+1$ spinon modes coincides with the result obtained from
(\ref{spinonS}) using the fusion rules of the CFT.  Furthermore we
have checked numerically for spins up to $S=3$ that the distribution
function (\ref{distk}) gives the correct behaviour \cite{babu} of the
low temperature specific heat \cite{scho:up}.

\begin{figure}
\epsfxsize=0.6\textwidth
\begin{center}
\leavevmode
\epsfbox{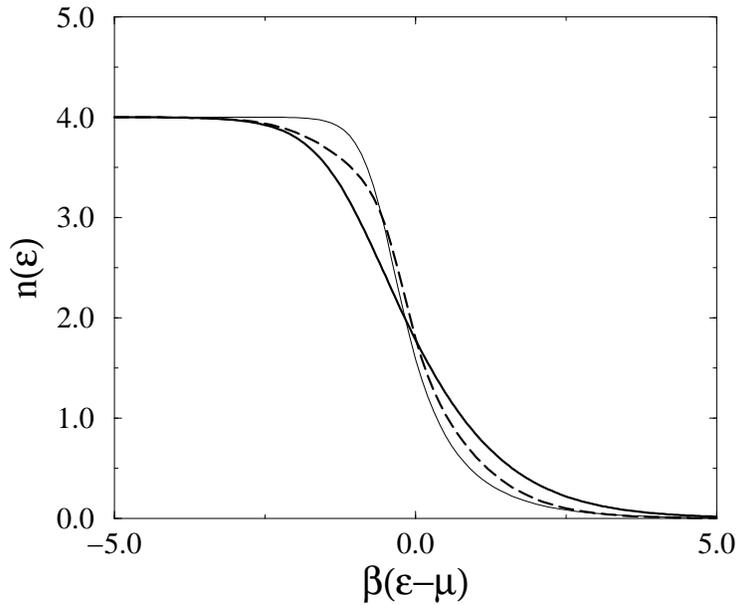}
\end{center}

\caption{Distribution function obtained from (\protect{\ref{distk}})
for the spinons of the spin-$1$ chain: shown are $n(\epsilon)$ for
vanishing magnetic field (bold curve) and for the fully polarized
spinon system (dashed curve) in comparison with the result for
$g={1\over4}$ exclusion statistics (thin curve).
\label{fig:nk}
}
\end{figure}

For $k=2$ the eigenvalues determining the distribution function
for $h=0$ and $h\to\infty$ (leading to complete polarization of the
spinons) are the largest real solutions of
\begin{equation}
  \begin{array}{ll}
   (\lambda-1)^2 = y^2(\lambda+1) &\, {\rm for~}h=0,\\[2pt]
   (\lambda-1)^2(\lambda+1) = y^2\ {\rm e}^{\beta h/2} \lambda^2 
	&\, {\rm for~}h\to\infty.
  \end{array}
\label{dist2}
\end{equation}
In Fig.~\ref{fig:nk} the distribution functions obtained from
(\ref{dist2}) are presented in comparison with the one for particles
obeying $g={1\over4}$ exclusion statistics.

To summarize, considering integrable lattice realizations of the
$SU(2)_k$ WZW models we have shown that the truncation scheme for
chiral spectra of a CFT introduced in \cite{scho:97} can be
interpreted as a recursion relation for partition functions of the
corresponding lattice model.  The truncated partition functions of the
field theory show complete equivalence between the well established spinon
picture for the integrable spin chain and the spinon formulation of
the CFT.  For the spin chains this method allows to compute
distribution functions for spinons obeying non-Abelian exchange
statistics.
Frequently, the low-lying excitations in one-dimensional $S={1\over2}$
systems have been interpreted by invoking a spinon-picture (see e.g.\
Ref.~\cite{SPINON}).  Similarly, we expect that the characterization of
the spinons in higher $S$ models given here will prove useful for a
better understanding of massive perturbations of the latter such as
the $S=1$ Haldane system.

The authors thank K.~Schoutens and J.~Fuchs for discussions.  This
work has been supported in parts by the Deutsche
For\-schungs\-gemeinschaft under Grant No.\ Fr~737/2--2.

\setlength{\baselineskip}{13pt}

\vfill

\begin{table}
\caption{String configurations for a Bethe state with given number $m$ of
  holes in the distribution of $2S$ strings and total spin $S_{tot}$:
  $[n^1p^2]$ denotes a solution with $\nu_n=1$, $\nu_p=2$.  For $S=2$ the
  configuration of strings with $n>2S$ ($<2S$) are listed separately, for a
  Bethe state the $n>2S$ configuration has to be supplemented by one of the
  possible $n<2S$ configurations listed in the last column.}
\begin{tabular}{cc|c|cc}
$m$ & $S_{tot}$ & $S={1\over2}$ & \multicolumn{2}{c}{$S=2$} \\
\tableline
2&1& --- &		--- & $[3^1]$ \\
 &0& $[2^1]$ & 		$[5^1]$ & $[3^1]$ \\
\tableline
4&2& --- &		--- & $[3^2]$,$[2^1]$ \\
 &1& $[2^1]$ &		$[5^1]$ & $[3^2]$,$[2^1]$ \\
 &0& $[3^1]$,$[2^2]$ &	$[6^1]$,$[5^2]$ & $[3^2]$,$[2^1]$ \\
\tableline
6&3& --- &		--- & $[3^3]$,$[3^1 2^1]$,$[1^1]$ \\
 &2& $[2^1]$ &		$[5^1]$ & $[3^3]$,$[3^1 2^1]$,$[1^1]$ \\
 &1& $[3^1]$,$[2^2]$ &	$[6^1]$,$[5^2]$ & $[3^3]$,$[3^1 2^1]$,$[1^1]$ \\
 &0& $[4^1]$,$[3^1 2^1]$,$[2^3]$ &
			$[7^1]$,$[6^1 5^1]$,$[5^3]$ & 
				$[3^3]$,$[3^1 2^1]$,$[1^1]$
\end{tabular}
\label{tab:no_conf}
\end{table}


\end{document}